\documentclass{article}
\usepackage[T1]{fontenc}
\usepackage[utf8]{inputenc}
\newif\ifarxivpreprint
\arxivpreprinttrue
\ifarxivpreprint
  \usepackage{ismir}
\else
  \usepackage[submission]{ismir}
\fi
\usepackage{amsmath,cite,url}
\usepackage{graphicx}
\usepackage{color}
\usepackage{booktabs}
\usepackage{xcolor}
\usepackage{colortbl}
\usepackage{multirow}
\ifarxivpreprint
  \usepackage{arxiv-theme}
  \definecolor{LightBluePurple}{HTML}{E7F8F5}
\else
  \definecolor{LightBluePurple}{HTML}{E8E8FF}
\fi
\newcommand{\zhuolabmark}{\raisebox{0.35ex}{\includegraphics[height=0.8em]{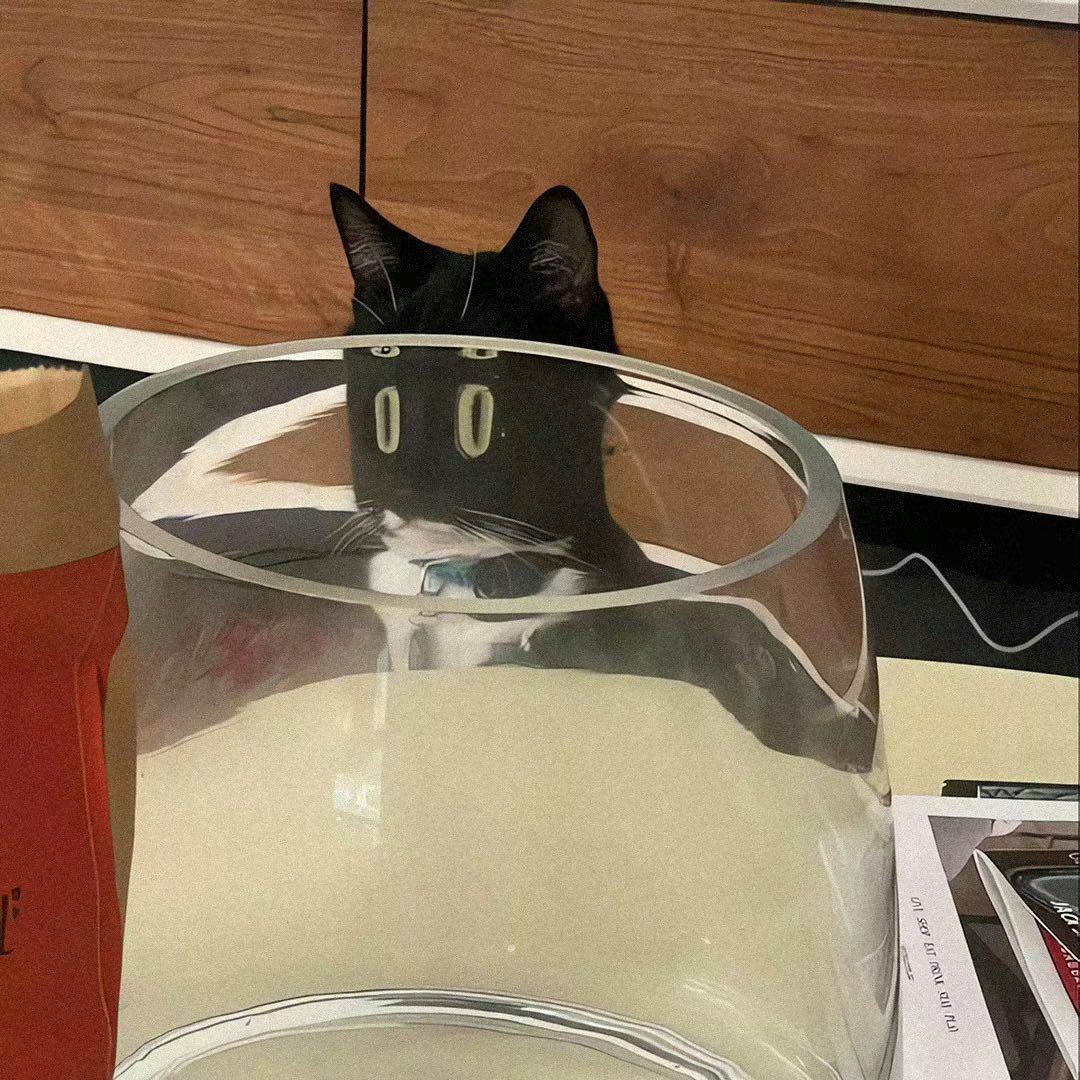}}}
\title{Real-Time Interactive Music Generation via Data-Free Streaming Consistency Distillation}

\threeauthors
  {Baisen Wang*\zhuolabmark} {DynamiX\\\texttt{wbs2788@gmail.com}}
  {Chenxi Bao*\zhuolabmark} {DynamiX}
  {Qisong Han*} {Cardiff University}

\begin{document}

\maketitle
\begingroup
  \renewcommand\thefootnote{*}
  \footnotetext{Equal contribution.}
\endgroup
\begingroup
  \renewcommand\thefootnote{\raisebox{-0.35ex}{\zhuolabmark}}
  \footnotetext{Work done in ZhuoLab.}
\endgroup
\begin{abstract}
Interactive music and live performance relies on real-time human expression, but modern generative music AI remains largely absent from this domain due to its prohibitive inference latency and offline rendering paradigm. To provide pioneer musicians with a novel medium for interactive composition, we should fundamentally change these static models into dynamic, playable instruments. In this paper, we propose a framework that bridges this gap. To achieve the low latency required for live interaction without sacrificing structural coherence, we formulate distillation within a streaming autoregressive latent space. Our approach gets rid of the need for expensive paired audio-latent datasets by utilizing prompt-only inputs to synthesize teacher-guided, chunk-wise trajectories on the fly. Because live instruments require high acoustic fidelity, we introduce music-aware consistency objectives, which combine latent, spectral, and temporal-difference losses, to preserve crucial qualities like timbre, transients, and rhythmic stability during accelerated single-step streaming generation. Implemented via parameter-efficient adaptation, our distillation reduces generation steps to achieve a low real-time factor. Crucially, by operating as a continuous autoregressive stream, the system can seamlessly assimilate dynamic human inputs on the fly, allowing users to instantly steer the musical trajectory without interrupting the audio flow. Ultimately, this work recontextualizes generative text-to-music models not as passive prompt-and-wait systems, but as responsive instruments, opening new frontiers for live human-AI musical co-creation.
\end{abstract}

\section{Introduction}\label{sec:introduction}

\begin{figure*}[t]
  \centering
  \includegraphics[width=\textwidth]{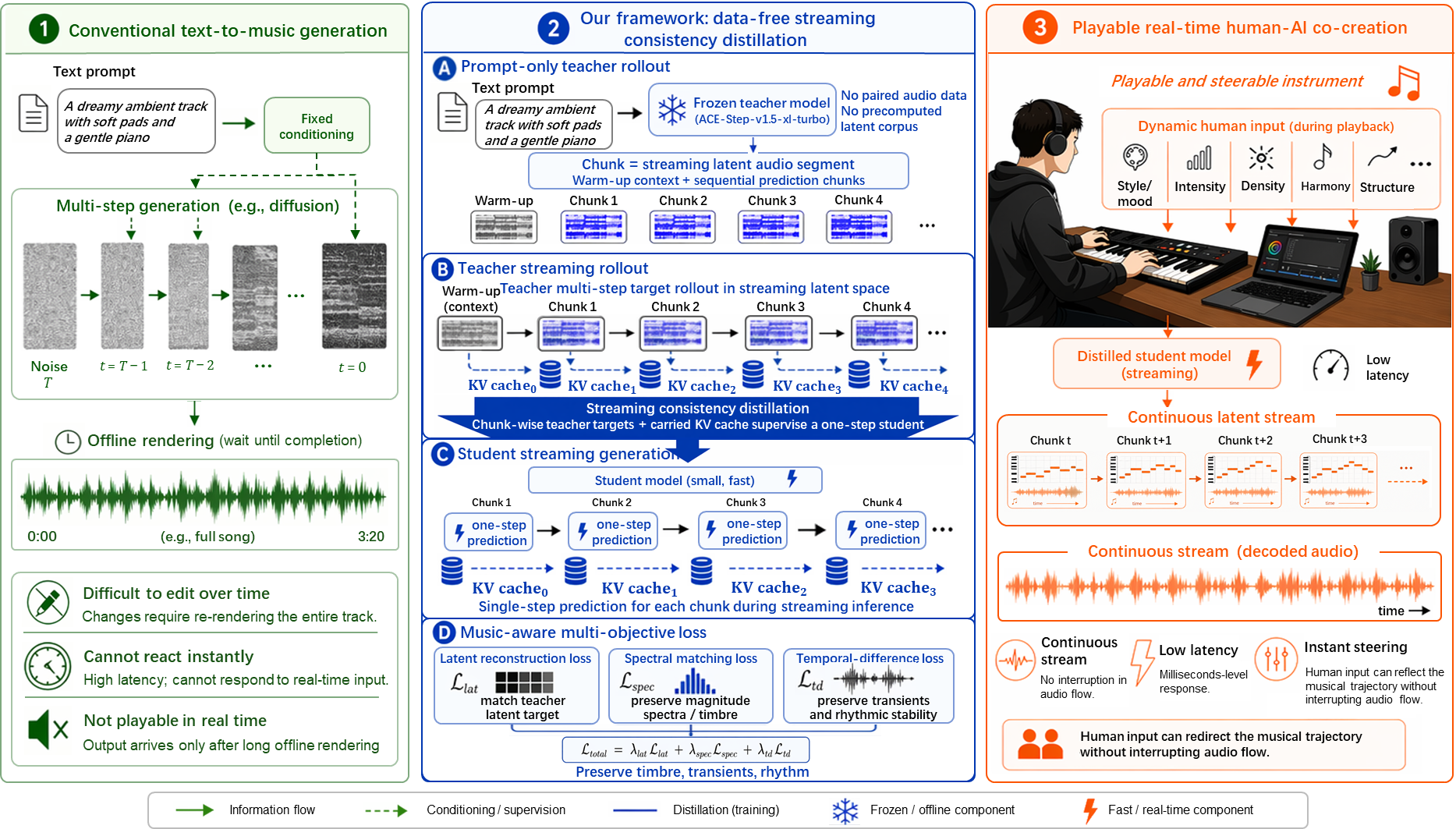}
  \caption{Overview of the proposed data-free streaming consistency distillation framework. Conventional text-to-music models rely on multi-step offline rendering, whereas our method synthesizes target chunks online through prompt-only teacher rollouts and trains a student with chunk-wise streaming consistency and music-aware objectives. The distilled student operates over a continuous latent stream, enabling low-latency human--AI co-creation with uninterrupted playback.}
  \label{fig:framework}
\end{figure*}

Musical expression unfolds through time and depends on continuous real-time interaction, yet most generative music models remain designed for static, offline generation. Although recent text-to-music systems like Suno, Stable Audio, and ACE-Step Series, can synthesize high-quality and structurally coherent audio, they still largely formulate music generation as one-shot rendering from fixed conditioning, in a manner more reminiscent of image synthesis than live musical behavior. This formulation makes it difficult to modify musical trajectories along the time axis or to incorporate new human input on the fly without restarting or disrupting the generation process. To address this gap, we introduce a data-free streaming consistency distillation framework that enables single-step autoregressive music generation for real-time human–AI co-creation.

This mismatch is not merely a matter of inference speed. A playable generative music system must continuously extend an evolving musical stream while remaining immediately responsive to new controls, all without sacrificing the perceptual qualities that make musical interaction meaningful. Achieving this is challenging because low-latency generation often comes at the cost of degraded timbral fidelity, weakened transient structure, unstable rhythm, or reduced long-range coherence, especially in streaming settings where errors can accumulate over time. More fundamentally, interactive music generation demands a model that treats music not as a fixed object to be rendered once, but as a temporally unfolding process that can be steered, revised, and continued seamlessly during playback.

To address these challenges, we formulate music generation as a streaming autoregressive process in latent space, where audio is produced chunk by chunk and can be continuously steered over time. Within this sequential generation framework, we use consistency distillation to compress the teacher’s multi-step prediction for each chunk into a low-latency student, ultimately enabling single-step generation during streaming inference. To avoid the cost of paired audio training data, the teacher synthesizes chunk-wise target trajectories online from prompt-only inputs, allowing distillation to proceed without precomputed audio-latent corpora. Because aggressive acceleration can easily damage perceptually important musical attributes, we further introduce music-aware consistency objectives that jointly constrain latent reconstruction, spectral structure, and temporal variation, helping preserve timbral fidelity, transient detail, and rhythmic stability. In this way, the distilled model is not merely faster, but capable of functioning as a continuously playable and steerable instrument for real-time human–AI musical interaction.

Our contributions are fourfold:
\begin{itemize}
    \setlength{\itemsep}{0pt}
    \setlength{\parsep}{0pt}
    \setlength{\topsep}{2pt}
    \item We propose data-free streaming consistency distillation for long-form text-to-music generation with prompt-only online supervision.
    \item We distill in a streaming autoregressive latent space with chunk-wise cached context for low-latency, extended-context generation.
    \item We introduce music-aware latent, spectral, and temporal-difference objectives to preserve timbre, transients, and rhythmic stability under extreme step reduction.
    \item We demonstrate real-time human--AI interaction, reframing the model as a phrase-level, semantically steerable generative instrument.
\end{itemize}

\section{Related Work}\label{sec:related_work}

\textbf{Text-to-music generation} has developed through diffusion-based and autoregressive paradigms. Diffusion and latent diffusion models \cite{ddpm, latent_diffusion} have been adapted to music or audio synthesis in Noise2Music, AudioLDM 2, Mo\^usai, Stable Audio, and ACE-Step \cite{noise2music, audioldm, mousai, stableaudio, gong2026acestep}, while autoregressive systems such as MusicGen, HeartMuLa, and SongGen use codec-token or transformer-based sequential modeling for controllability and structure \cite{musicgen, yang2026heartmula, liu2025songgen}. Most of these models target complete-sequence generation rather than low-latency playback. Streaming generation has been explored in StreamFlow \cite{choistreamflow}, with related chunk-wise or causal formulations in singing voice, speech, and motion generation \cite{cui2025cssinger, li2026robust, xiao2025motionstreamer}; our method brings this perspective into latent-space text-to-music generation.

\noindent\textbf{Diffusion acceleration} is commonly achieved with fast samplers or distillation \cite{lu2022dpm, salimans2022progressive}. Consistency models learn mappings from noisy trajectory points to clean samples for few-step or single-step generation \cite{song2023consistency}, and audio variants such as ConsistencyTTA and MusicCM apply this idea to text-to-audio and music synthesis \cite{bai2023consistencytta, fei2024music}. Because extreme acceleration can compromise acoustic fidelity, we further incorporate spectral and temporal-domain constraints motivated by audio generation and vocoding studies \cite{park2025bemaganv2}.

\noindent\textbf{Interactive AI music systems} increasingly emphasize human-in-the-loop control, including preference-based search \cite{marcos2025random}, gesture- or expression-driven steering \cite{mamotion}, and high-level or time-varying musical controls \cite{tan2020music, music_controlnet}. Unlike these works, which mainly address interfaces, control representations, or post-generation search, our framework targets the generative backbone by converting an offline renderer into a low-latency continuous stream.

\section{Method}\label{sec:method}

\begin{figure*}[t]
  \centering
  \includegraphics[width=\textwidth]{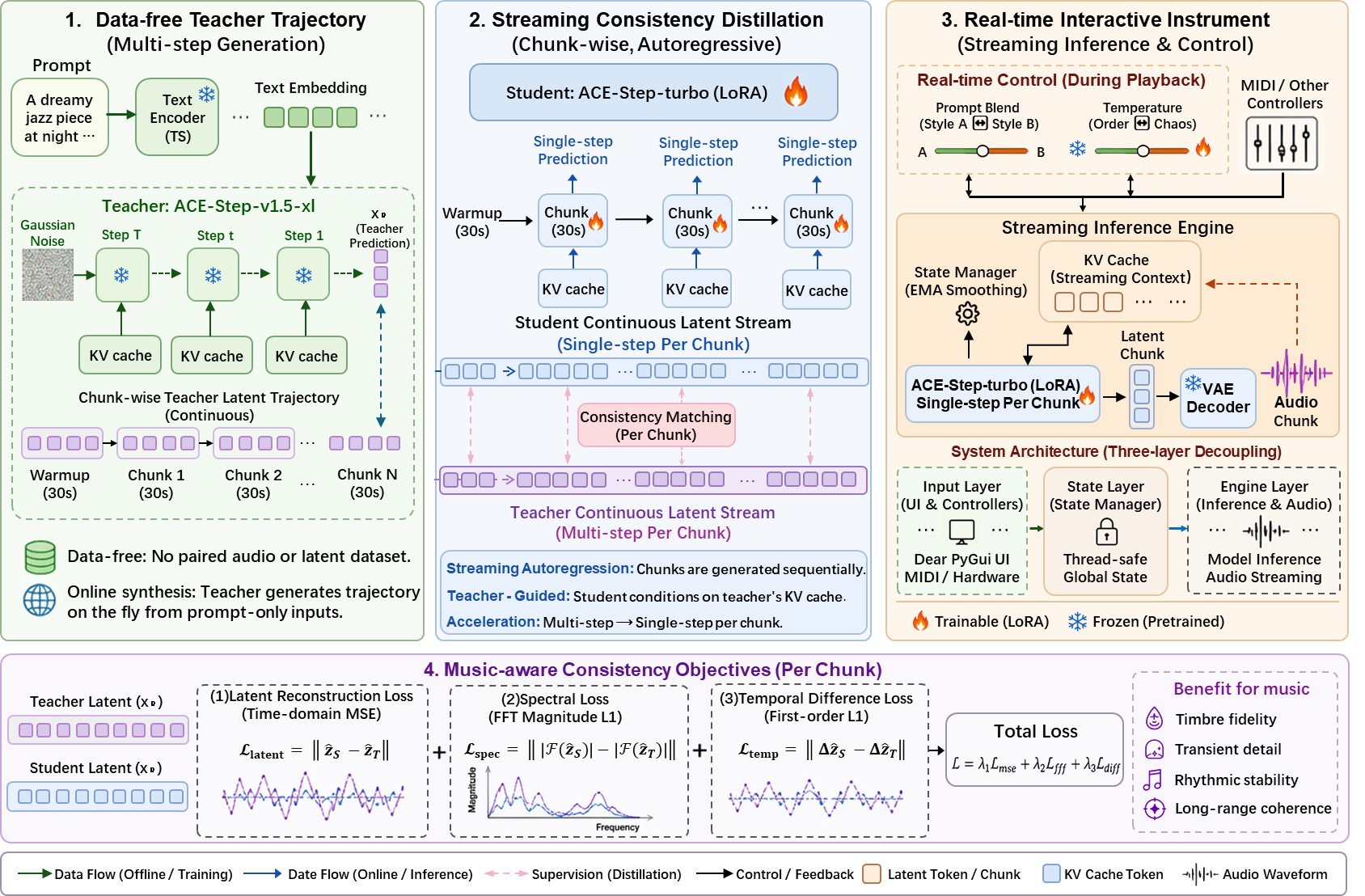}
  \caption{Detailed illustration of the proposed data-free streaming consistency distillation method. (1) The frozen teacher generates chunk-wise latent trajectories via multi-step sampling. (2) The student learns to predict the same latent chunk in a single step, conditioned on the teacher's streaming context (KV cache). (3) The decoupled system architecture supports real-time interactive generation. (4) Music-aware consistency objectives combine latent reconstruction, spectral matching, and temporal-difference losses to preserve acoustic fidelity.}
  \label{fig:method_details}
\end{figure*}

\subsection{Problem Formulation}

We formulate long-form text-to-music generation as a streaming process in latent space. Given a text prompt $p$, a continuous latent trajectory $\mathcal{Z} = [z^{(0)}, z^{(1)}, \dots, z^{(K)}]$ is partitioned into an initial warm-up segment $z^{(0)}$ and a sequence of autoregressive prediction chunks $z^{(k)}$ for $k \geq 1$.

Our distillation framework operates in a \emph{data-free} regime with respect to paired audio--latent supervision: rather than retrieving ground-truth audio latents, supervision is synthesized online by rolling out a frozen teacher model. At chunk $k$, the teacher $f_T$ carries forward a historical context state $c_T^{(k-1)}$, which contains the streaming context accumulated from previous chunks, such as transformer KV-cache states. Starting from Gaussian noise $\epsilon \sim \mathcal{N}(0, \mathbf{I})$, the teacher executes an $N$-step ODE solver to produce a target latent chunk:
\begin{equation}
\hat{z}_T^{(k)} = f_T\left(\epsilon, p, c_T^{(k-1)}; N\right).
\end{equation}
The student model $f_S$ is trained to reproduce the same target using a much smaller sampling budget of $M$ steps, where $M \ll N$ and typically $M=1$:
\begin{equation}
\hat{z}_S^{(k)} = f_S\left(\epsilon, p, c_T^{(k-1)}; M\right).
\end{equation}

Conditioning the student on the teacher-generated streaming context exposes it to long-context dependencies that arise during sequential generation. This teacher-guided formulation encourages cross-chunk continuity and mitigates boundary artifacts during streaming inference. The overall objective is to minimize the chunk-wise discrepancy between teacher and student predictions:
\begin{equation}
\min_{\theta_S} \sum_{k=1}^{K} \mathcal{L}\left(\hat{z}_T^{(k)}, \hat{z}_S^{(k)}\right),
\end{equation}
where $\mathcal{L}$ denotes the proposed music-aware consistency objective. This formulation unifies prompt-only supervision, streaming context propagation, and extreme sampling acceleration within a single training framework.

\subsection{Data-Free Teacher Rollout via Online Synthesis}

A central goal of our framework is to eliminate reliance on paired audio data and precomputed latent trajectories. We therefore introduce a \emph{prompt-only} training regime in which supervision is synthesized online by a frozen teacher model. Given a text prompt $p$, we compute its conditioning embeddings and initialize a streaming generation process directly in latent space. Instead of retrieving ground-truth audio latents from a static dataset, the teacher constructs the supervision trajectory through its own rollout.

For each training instance, the teacher first synthesizes a warm-up segment $z^{(0)}$ to establish the contextual state $c_T^{(0)}$ (i.e., the KV-cache) required for long-form continuation. It then proceeds autoregressively, predicting a sequence of target chunks:
\begin{equation}
\hat{z}_T^{(1)}, \hat{z}_T^{(2)}, \dots, \hat{z}_T^{(K)}.
\end{equation}
Each chunk $\hat{z}_T^{(k)}$ is conditioned on both the text prompt $p$ and the accumulated streaming context $c_T^{(k-1)}$. The rollout yields a coherent sequential latent trajectory:
\begin{equation}
\hat{\mathcal{Z}}_T = [z^{(0)}, \hat{z}_T^{(1)}, \hat{z}_T^{(2)}, \dots, \hat{z}_T^{(K)}].
\end{equation}
This trajectory matches the autoregressive streaming protocol used by the student during inference.

Generating teacher rollouts \emph{online}, rather than caching them as a static corpus, has two advantages. First, it avoids the storage and preprocessing bottlenecks associated with paired audio--latent datasets, which become especially costly for long-form music. Second, because the teacher starts from newly sampled noise during training, the student observes diverse valid trajectories for the same prompt, providing implicit regularization. For every prediction chunk $k$, the teacher outputs both the target latent $\hat{z}_T^{(k)}$ and the updated context $c_T^{(k)}$ required for the next step. Prompt-only online synthesis therefore provides dynamic supervision while preserving the temporal structure required by streaming music generation.

\subsection{Streaming Consistency Distillation via Chunk-Wise Cached Trajectories}

Standard consistency distillation typically optimizes isolated segments. When applied directly to long-form audio, this can introduce discontinuities at segment boundaries. To preserve cross-chunk coherence, we distill over the teacher's \emph{chunk-wise cached trajectories}, defined as $\left\{(c_T^{(k-1)}, \hat{z}_T^{(k)})\right\}_{k=1}^{K}$. By explicitly pairing each target chunk $\hat{z}_T^{(k)}$ with its corresponding historical KV-cache state $c_T^{(k-1)}$, the student learns from the context evolution of a continuous streaming process.

As detailed in the center panels of \figref{fig:method_details}, for each prediction step $k$, the student is conditioned on the teacher-provided context to predict the target in $M$ steps ($M \ll N$):
\begin{equation}
\hat{z}_S^{(k)} = f_S\left(\epsilon, p, c_T^{(k-1)}; M\right).
\end{equation}
This teacher-guided mechanism is important for music generation. Rather than autoregressing on its own potentially imperfect history during training, the student remains anchored to the teacher trajectory. This reduces the risk that local prediction errors cascade into rhythmic drift, timbral instability, or audible boundary artifacts. The formulation compresses the multi-step generative process to as few as $M=1$ step per chunk, enabling low-latency streaming inference.

\subsection{Music-Aware Consistency Objectives}

Standard consistency objectives often minimize only latent distance. Under extreme step reduction, however, pure latent MSE may over-smooth the prediction, leading to muffled timbres and smeared rhythmic transients. We therefore augment the basic latent distance with spectral and temporal-difference penalties that operate directly on the latent embeddings. This design encourages musically relevant structure while avoiding the memory and compute overhead of VAE decoding during training.

For each chunk $k$, let $\hat{z}_T^{(k)}$ and $\hat{z}_S^{(k)}$ denote the teacher and student latent predictions, respectively. We first define the base latent reconstruction loss:
\begin{equation}
\mathcal{L}_{\mathrm{latent}}^{(k)} = \left\| \hat{z}_S^{(k)} - \hat{z}_T^{(k)} \right\|_2^2,
\end{equation}
which anchors the student to the teacher's chunk-level trajectory.

To retain frequency-domain fidelity, we apply a one-dimensional real fast Fourier transform (RFFT) along the temporal dimension of the latents. The spectral consistency loss penalizes the L1 distance between magnitude spectra:
\begin{equation}
\mathcal{L}_{\mathrm{spec}}^{(k)} = \left\|\, |\mathcal{F}(\hat{z}_S^{(k)})| - |\mathcal{F}(\hat{z}_T^{(k)})| \,\right\|_1,
\end{equation}
where $\mathcal{F}(\cdot)$ denotes the Fourier transform and $|\cdot|$ denotes magnitude. Matching the latent spectral energy distribution encourages the student to preserve the teacher's acoustic texture and timbral characteristics.

Finally, to preserve structural sharpness, including percussive transients and rhythmic boundaries, we apply a first-order temporal-difference operator $\Delta$ across adjacent latent frames:
\begin{equation}
\mathcal{L}_{\mathrm{temp}}^{(k)} = \left\| \Delta \hat{z}_S^{(k)} - \Delta \hat{z}_T^{(k)} \right\|_1.
\end{equation}
We use an L1 penalty to encourage sparse derivative errors, which supports localized transient alignment while discouraging diffuse artifacts introduced by aggressive acceleration.

The full music-aware consistency objective is defined as:
\begin{equation}
\mathcal{L}^{(k)} = \lambda_{\mathrm{latent}} \mathcal{L}_{\mathrm{latent}}^{(k)} + \lambda_{\mathrm{spec}} \mathcal{L}_{\mathrm{spec}}^{(k)} + \lambda_{\mathrm{temp}} \mathcal{L}_{\mathrm{temp}}^{(k)},
\end{equation}
resulting in the final streaming distillation loss $\mathcal{L} = \sum_{k=1}^{K} \mathcal{L}^{(k)}$. Together, these objectives encourage the student to track the teacher's latent predictions while preserving the timbral consistency and rhythmic stability required for high-fidelity streaming generation.

\begin{table}[t]
\centering

{\small
\setlength{\tabcolsep}{4pt}
\renewcommand{\arraystretch}{0.92}
\begin{tabular}{@{}lcccc@{}}
\toprule
\textbf{Method} & \textbf{Steps} & \textbf{Stream} & \textbf{TTFA $\downarrow$} & \textbf{RTF $\downarrow$} \\
\midrule
Original  & 8 & No  & 0.708 & 0.024 \\
Distilled (forced) & 8 & No  & 1.213 & 0.040 \\
Distilled & 1 & No  & 1.148 & 0.038 \\
\rowcolor{LightBluePurple}
\textbf{Ours} & \textbf{1} & \textbf{Yes} & \textbf{0.086} & \textbf{0.009} \\
\bottomrule
\end{tabular}
}
\caption{Latency comparison under the benchmark protocol. Original denotes ACE-Step XL-Turbo; Distilled denotes the one-step consistency LoRA student. Lower is better.}\label{tab:latency_paradigm}
\end{table}

\begin{table}[b]
\centering

{\small
\setlength{\tabcolsep}{4pt}
\renewcommand{\arraystretch}{0.92}
\begin{tabular}{@{}lccc@{}}
\toprule
\textbf{$C_{\mathrm{sec}}$} & \textbf{TTFA $\downarrow$} & \textbf{RTF $\downarrow$} & \textbf{Control Latency $\downarrow$} \\
\midrule
0.5 & 0.085 & 0.009 & 0.543 \\
1.0 & 0.083 & 0.008 & 1.042 \\
1.5 & 0.085 & 0.008 & 1.543 \\
2.0 & 0.084 & 0.008 & 2.043 \\
\bottomrule
\end{tabular}
}
\caption{Streaming chunk-size ablation. Control latency is estimated from the online scheduling model. Lower is better.}\label{tab:chunk_ablation}
\end{table}

\subsection{Real-Time Streaming Inference and Interactive Control}

At inference time, the distilled student mirrors the training protocol in a chunk-wise streaming regime. Given a text prompt $p$, generation begins with an initial warm-up chunk of $T_{\mathrm{warm}}$ latent frames, followed by an open-ended sequence of prediction chunks of $T_{\mathrm{pred}}$ frames. The student maintains its own KV-cache state across chunk boundaries, avoiding the need to recompute historical attention.

For each prediction chunk $k$, the model receives a fresh Gaussian latent initialization $z_t^{(k)}$ (typically sampled at $t=1$), the text-conditioned encoder states, and the KV-cache propagated from chunk $k-1$. Under the velocity parameterization, a single decoder forward pass predicts $v_{\mathrm{pred}}^{(k)}$, from which the clean latent chunk is recovered via a one-step Euler update:
\begin{equation}
\hat{z}_S^{(k)} = z_t^{(k)} - t \cdot v_{\mathrm{pred}}^{(k)}.
\end{equation}
The updated cache is propagated to chunk $k+1$, enabling continuous long-context generation. Each predicted latent chunk $\hat{z}_S^{(k)}$ is then asynchronously decoded into waveform audio by the VAE. Because the student uses only $M=1$ step per chunk, latent generation can outpace real-time playback and yield low startup latency.

\subsection{Interactive Instrument Paradigm}

The proposed system is designed to be played as a phrase-level generative instrument. Its playability does not arise from note-level actuation, as in a keyboard or drum controller, but from continuous expressive control over an unfolding musical process. During performance, the current audio chunk is played while the next chunk is generated in the background, so performer actions shape forthcoming material without stopping or regenerating the full track.

The control vocabulary is semantic rather than symbolic. Performers can trigger scene-level states such as \emph{intro}, \emph{build}, \emph{breakdown}, or \emph{outro}; adjust musical dimensions such as energy, density, brightness, tension, and rhythmic drive; or steer arrangement by adding or reducing instrumental layers. These controls are compiled into conditioning-state updates, including prompt interpolation, semantic prompt modifiers, and sampling-temperature adjustment. Because the autoregressive cache is preserved across chunks, interventions bend the continuation from the current trajectory rather than resetting it. Thus, the system functions as a live semantic instrument for shaping structure, timbre, density, and direction, while RTF and control latency respectively characterize playback throughput and the audibility of performer intent.

\begin{table}[t]
\centering
{\small
\setlength{\tabcolsep}{3pt}
\renewcommand{\arraystretch}{0.92}
\begin{tabular}{@{}lcccccc@{}}
\toprule
\textbf{Mode} & \textbf{Loss} & \textbf{$K$} & \textbf{$S$} & \textbf{CLAP $\uparrow$} & \textbf{KLD $\downarrow$} & \textbf{FD $\downarrow$} \\
\midrule
\multirow{12}{*}{Streaming}
& $\mathcal L_{\mathrm{latent}}$ & 1 & 1 & 0.329 & 0.693 & 304.83 \\
& $+\mathcal L_{\mathrm{spec}}$ & 1 & 1 & 0.348 & 0.655 & 298.73 \\
& $+\mathcal L_{\mathrm{temp}}$ & 1 & 1 & 0.344 & 0.705 & 295.96 \\
& $\mathcal L_{\mathrm{full}}$ & 1 & 1 & 0.361 & 0.635 & 294.66 \\
\cmidrule{2-7}
& $\mathcal L_{\mathrm{latent}}$ & 3 & 1 & 0.257 & 0.324 & 398.79 \\
& $+\mathcal L_{\mathrm{spec}}$ & 3 & 1 & 0.294 & 0.322 & 388.51 \\
& $+\mathcal L_{\mathrm{temp}}$ & 3 & 1 & 0.278 & 0.319 & 378.75 \\
& $\mathcal L_{\mathrm{full}}$ & 3 & 1 & 0.300 & 0.302 & 349.32 \\
\cmidrule{2-7}
& $\mathcal L_{\mathrm{latent}}$ & 5 & 1 & 0.262 & 0.307 & 389.13 \\
& $+\mathcal L_{\mathrm{spec}}$ & 5 & 1 & 0.273 & 0.308 & 389.59 \\
& $+\mathcal L_{\mathrm{temp}}$ & 5 & 1 & 0.275 & 0.317 & 378.17 \\
& $\mathcal L_{\mathrm{full}}$ & 5 & 1 & 0.282 & 0.308 & 367.22 \\
\midrule
\multirow{12}{*}{Non-stream}
& $\mathcal L_{\mathrm{latent}}$  & 1 & 1 & 0.319 & 0.740 & 308.99 \\
& $\mathcal L_{\mathrm{latent}}$ & 1 & 4 & 0.320 & 0.678 & 303.17 \\
& $\mathcal L_{\mathrm{latent}}$ & 1 & 8 & 0.320 & 0.722 & 308.43 \\
\cmidrule{2-7}
& $+\mathcal L_{\mathrm{spec}}$ & 1 & 1 & 0.348 & 0.655 & 298.73 \\
& $+\mathcal L_{\mathrm{spec}}$ & 1 & 4 & 0.351 & 0.613 & 297.73 \\
& $+\mathcal L_{\mathrm{spec}}$ & 1 & 8 & 0.347 & 0.616 & 292.20 \\
\cmidrule{2-7}
& $+\mathcal L_{\mathrm{temp}}$ & 1 & 1 & 0.348 & 0.665 & 299.36 \\
& $+\mathcal L_{\mathrm{temp}}$ & 1 & 4 & 0.336 & 0.674 & 304.80 \\
& $+\mathcal L_{\mathrm{temp}}$ & 1 & 8 & 0.338 & 0.699 & 303.09 \\
\cmidrule{2-7}
& $\mathcal L_{\mathrm{full}}$ & 1 & 1 & 0.361 & 0.635 & 294.66 \\
& $\mathcal L_{\mathrm{full}}$ & 1 & 4 & 0.358 & 0.660 & 296.55 \\
& $\mathcal L_{\mathrm{full}}$ & 1 & 8 & 0.354 & 0.633 & 293.69 \\
\bottomrule
\end{tabular}
}
\caption{Objective quality evaluation across objectives and inference modes. $K$ denotes evaluated prediction chunks and $S$ denotes denoising steps; KLD and FD denote PaSST-KLD and OpenL3-FD, respectively.}
\label{tab:stable_audio_comprehensive}
\end{table}

\section{Experiments}\label{sec:experiments}

\subsection{Experimental Setup}\label{subsec:setup}

We build on ACE-Step 1.5 XL-Turbo \cite{gong2026acestep}. The teacher remains frozen, and the student uses the same backbone with LoRA adaptation \cite{TL:LoRA} on the DiT decoder \cite{dit}. The LoRA adapter is applied to the query, key, value, and output projections with rank 64, scaling factor 128, and dropout 0.1. Training uses bfloat16 (bf16) precision, batch size 32, gradient accumulation 1, AdamW \cite{adamw}, cosine learning-rate decay, initial learning rate \(1\times10^{-4}\), 100 warm-up steps, weight decay 0.01, gradient clipping at 1.0, seed 42, and 2,000 optimization steps.

For each training sample, the teacher synthesizes a streaming trajectory with a 30-second warm-up segment followed by up to five 30-second prediction chunks. This setting yields a 150-second prediction horizon after warm-up and a 180-second generated trajectory during distillation. Training uses prompt-only supervision from a pool of 125,446 natural-language music descriptions, rather than paired target audio. Unless otherwise stated, $\lambda_{\mathrm{latent}}=1$, and the auxiliary loss weights are binary according to the ablation: $\mathcal{L}_{\mathrm{latent}}$, $+\mathcal{L}_{\mathrm{spec}}$, $+\mathcal{L}_{\mathrm{temp}}$, or $\mathcal{L}_{\mathrm{full}}$.

Latency is evaluated on a single NVIDIA H200 GPU with batch size 1, using a filtered SongDescriber benchmark \cite{songdescriber} containing 279 instrumental items. We sample 10 prompts and repeat each prompt three times, yielding 30 runs per condition. All prompts use the \texttt{caption} field and lyrics are fixed to \texttt{[Instrumental]}. We follow the benchmark timing protocol: standard latency is measured as the wall-clock time of one \texttt{generate\_music} call without disk I/O, while streaming startup latency measures the time to the first streaming rollout; RTF is computed over the full generated duration and includes decoding in the streaming setting. Control latency is estimated from the online scheduling model as the chunk duration plus the average prediction-chunk latency.

\subsection{Generation Efficiency and Interactivity}\label{subsec:efficiency}

Table~\ref{tab:latency_paradigm} separates sampling acceleration from streaming reformulation. The consistency-trained student is optimized for one-step prediction; applying it in an 8-step non-streaming sampler therefore serves only as a stress-test baseline rather than as the intended inference mode. One-step non-streaming inference reduces the sampling budget but still waits for a full batch output before playback. The streaming formulation reaches 0.086\,s startup latency and achieves the lowest RTF, indicating that chunk-wise generation is the primary factor behind the low interactive latency. All rows use the same prompt set and checkpoint family, while the streaming row uses the chunk-wise inference path required for online playback.

Table~\ref{tab:chunk_ablation} varies the prediction chunk duration $C_{\mathrm{sec}}$. TTFA and RTF remain nearly unchanged from 0.5 to 2.0\,s, while estimated control latency follows the expected chunk-boundary scheduling cost. Thus, $C_{\mathrm{sec}}$ directly controls the responsiveness--scheduling tradeoff; we use 1.0\,s as the default setting.

\begin{table}[t]
\centering
{\small
\setlength{\tabcolsep}{4pt}
\renewcommand{\arraystretch}{0.92}
\begin{tabular}{@{}cccc@{}}
\toprule
\textbf{$C_{\mathrm{sec}}$} & \textbf{CLAP $\uparrow$} & \textbf{KLD $\downarrow$} & \textbf{FD $\downarrow$} \\
\midrule
0.5 & 0.256 & 0.352 & 291.91 \\
1.0 & 0.268 & 0.327 & 283.71 \\
1.5 & 0.292 & 0.307 & 271.70 \\
2.0 & 0.274 & 0.302 & 277.41 \\
\bottomrule
\end{tabular}
}
\caption{Objective quality under different streaming chunk durations. All rows use one-step streaming with $\mathcal{L}_{\mathrm{full}}$.}
\label{tab:chunk_quality}
\end{table}

\begin{table*}[b]
\centering
{\small
\setlength{\tabcolsep}{3pt}
\renewcommand{\arraystretch}{0.92}
\begin{tabular}{@{}lccccc@{}}
\toprule
\textbf{Configuration} & \textbf{O-MOS $\uparrow$} & \textbf{R-MOS $\uparrow$} & \textbf{Resp. $\uparrow$} & \textbf{Steer. $\uparrow$} & \textbf{Co-create $\uparrow$} \\
\midrule
Ground Truth & $\mathbf{4.66} \pm 0.09$ & $\mathbf{4.70} \pm 0.09$ & N/A & N/A & N/A \\
Teacher Offline & $4.18 \pm 0.08$ & $4.24 \pm 0.09$ & $1.82 \pm 0.12$ & $2.35 \pm 0.13$ & $2.12 \pm 0.13$ \\
Non-stream ($S=1$) & $3.86 \pm 0.09$ & $3.90 \pm 0.09$ & $2.18 \pm 0.12$ & $2.50 \pm 0.13$ & $2.42 \pm 0.13$ \\
Stream Latent & $3.54 \pm 0.11$ & $3.62 \pm 0.10$ & $4.52 \pm 0.10$ & $3.88 \pm 0.12$ & $3.95 \pm 0.12$ \\
Stream +Spec & $3.74 \pm 0.10$ & $3.79 \pm 0.10$ & $4.50 \pm 0.10$ & $4.03 \pm 0.11$ & $4.10 \pm 0.11$ \\
Stream +Temp & $3.68 \pm 0.10$ & $3.73 \pm 0.10$ & $4.50 \pm 0.10$ & $4.08 \pm 0.11$ & $4.13 \pm 0.11$ \\
\rowcolor{LightBluePurple}
\textbf{Ours Full} & $3.92 \pm 0.09$ & $4.02 \pm 0.09$ & $\mathbf{4.55} \pm 0.10$ & $\mathbf{4.32} \pm 0.10$ & $\mathbf{4.38} \pm 0.10$ \\
\bottomrule
\end{tabular}
}
\caption{Subjective evaluation results reported as MOS $\pm$ 95\% confidence intervals. Ground-truth audio is evaluated only in passive listening.}
\label{tab:subjective_mos}
\end{table*}

\subsection{Objective Audio Quality Evaluation}\label{subsec:stable_audio_metrics}

We evaluate quality on the same SongDescriber-derived benchmark, using all 279 generated files per condition rather than the 30-run latency subset. We report CLAP score for text--audio alignment \cite{clap}, PaSST-KLD for music-tagging distributional divergence \cite{koutini2021efficient, music_tagging}, and OpenL3-FD for acoustic/perceptual fidelity \cite{cramer2019look}. The evaluation uses fixed preprocessing and metric settings, including peak normalization, CLAP at 48\,kHz, PaSST-KLD with 10\,s windows and 5\,s overlap, and OpenL3-FD over full-audio embeddings. Table~\ref{tab:stable_audio_comprehensive} compares loss objectives, rollout lengths, and denoising budgets under streaming and non-streaming inference.

The results show that the full music-aware objective improves the one-step streaming model. At $K=1$, the full objective outperforms the latent-only objective on all three metrics, increasing CLAP from 0.329 to 0.361 while reducing PaSST-KLD from 0.693 to 0.635 and OpenL3-FD from 304.83 to 294.66. The advantage is more pronounced over longer rollouts: at $K=3$ and $K=5$, the full objective yields the lowest OpenL3-FD among the streaming variants and the strongest CLAP score, suggesting better preservation of both prompt alignment and acoustic fidelity during autoregressive continuation. In the non-streaming setting, reducing $S$ from 8 to 1 does not cause severe degradation for the measured objectives, which supports the use of one-step inference as the computational basis for the streaming system. These quality results complement the latency analysis: step reduction preserves acceptable objective quality, while the streaming formulation is responsible for the reduced startup latency.

Table~\ref{tab:chunk_quality} evaluates whether the latency--responsiveness tradeoff in Table~\ref{tab:chunk_ablation} affects objective quality. Very short chunks ($C_{\mathrm{sec}}=0.5$) yield the weakest CLAP, PaSST-KLD, and OpenL3-FD scores, suggesting that overly frequent autoregressive boundaries can degrade musical continuity. Quality improves as the chunk duration increases to 1.5\,s, which achieves the best CLAP and OpenL3-FD, while 2.0\,s gives the best PaSST-KLD but slightly lower CLAP and OpenL3-FD. We therefore use 1.0\,s as a responsiveness-oriented default and view 1.5\,s as a quality-oriented operating point when additional control latency is acceptable.

\subsection{Subjective Evaluation}\label{subsec:user_study}

Because objective metrics do not capture interaction quality, we conduct a two-phase subjective evaluation with $N=20$ participants. Phase 1 uses a 5-point MOS listening test over five static audio clips per condition to evaluate overall quality (O-MOS) and prompt relevance (R-MOS). Phase 2 asks participants to steer a generated track between genres using an interactive interface, and rates responsiveness, steerability, and co-creation experience. We report MOS scores with 95\% confidence intervals. Ground-truth audio is included only in the passive listening phase because interaction metrics are not applicable to non-generative reference audio.

The subjective results are consistent with the quantitative latency and quality trends. Offline generation obtains relatively strong passive-listening scores but low interaction scores, reflecting the delay imposed by wait-and-listen rendering. In contrast, streaming variants substantially improve responsiveness, steerability, and co-creation scores. Among the streaming systems, the full objective achieves the strongest overall balance between passive quality and interactive usability.

\section{Conclusion}\label{sec:conclusion}

We presented a data-free streaming consistency distillation framework for real-time interactive music generation. The method synthesizes teacher trajectories from prompt-only inputs and trains a one-step student in a chunk-wise autoregressive latent space, enabling low-latency generation with cached streaming context, asynchronous decoding, and phrase-level user control. To preserve quality under strong acceleration, we introduced music-aware consistency objectives combining latent, spectral, and temporal-difference constraints.

Experiments show that streaming reformulation is key to low startup latency and real-time throughput, while the full objective improves text--audio alignment, acoustic fidelity, responsiveness, and steerability over offline baselines. Overall, this work moves text-to-music generation from prompt-and-wait rendering toward a semantically steerable generative instrument. Future work will reduce control-to-audio latency and explore richer interfaces for fine-grained musical interaction.

\bibliography{ISMIRtemplate}

%
%
%
%

\end{document}